\documentclass[11pt]{article}

\usepackage{graphicx}
\usepackage{amssymb,amsmath}
\usepackage{hyperref}

\newcommand{\de}{\mathrm{d}}
\newcommand{\ee}{\mathrm{e}}
\newcommand{\im}{\mathrm{i}}
\newcommand{\spc}{\,}
\newcommand{\qt}{\tilde{q}}
\newcommand{\pt}{\tilde{p}}

\newcommand{\Ec}{\mathcal{E}}
\newcommand{\Sc}{\mathcal{S}}

\newcommand{\VBH}{V_\mathrm{BH}}

\newcommand{\abs}[1]{\lvert #1\rvert}
\renewcommand{\Im}{\operatorname{Im}}
\renewcommand{\Re}{\operatorname{Re}}
\newcommand{\I}{\mathcal{I}}
\newcommand{\R}{\mathcal{R}}

\newcommand{\pd}{\partial}

\newcommand{\T}{^\mathrm{T}}
\newcommand{\bigO}{\mathcal{O}}

\numberwithin{equation}{section}

\begin{document}
\begin{titlepage}
\begin{center}
\hfill LMU-ASC 47/06\\
\hfill MPP-2006-88\\
\hfill {\tt hep-th/0607202}\\
\vskip 10mm

{\Large \textbf{Extremal non-BPS
black holes and entropy extremization }}
\vskip 8mm

\textbf{G.~L.~Cardoso$^{a}$,  V.~Grass$^{a}$,
D.~L\"ust$^{a,b}$ and J.~Perz$^{a,b}$}

\vskip 4mm
$^{a}${\em Arnold Sommerfeld Center for Theoretical Physics\\
Department f\"ur Physik,
Ludwig-Maximilians-Universit\"at M\"unchen \\
Theresienstra{\ss}e 37,
80333 M\"unchen, Germany}\\

\vskip 4mm
$^{b}${\em Max-Planck-Institut f\"ur Physik \\
F\"ohringer Ring 6,
80805 M\"unchen, Germany}\\
\vskip 4mm

{\tt gcardoso,viviane,luest,perz@theorie.physik.uni-muenchen.de}
\end{center}
\vskip .2in
\begin{center} {\bf ABSTRACT } \end{center}
\begin{quotation}\noindent
At the horizon, a static
extremal black hole solution
in $N=2$ supergravity
in four dimensions is determined by a set of so-called
attractor equations which, in the absence of higher-curvature interactions,
can be derived
as extremization conditions for the black hole potential or, equivalently, for the entropy function.
We contrast both methods by explicitly
solving the attractor equations for a one-modulus prepotential
associated with the conifold.  We find that near the conifold point,
the non-supersymmetric solution
has a substantially different behavior than the supersymmetric
solution.
We analyze the stability of the solutions and the extrema of the resulting entropy as a function of the modulus.
For the non-BPS solution the region of attractivity and the maximum of the
entropy do not coincide with the conifold point.

\end{quotation}

\vfill
\end{titlepage}

\eject

\section{Introduction}

In the near-horizon limit, static extremal black hole solutions in four dimensions are entirely determined in terms of the electric and magnetic charges carried by the black hole. In particular, the scalar fields at the horizon take values governed by a set of so-called attractor equations, involving these charges.  The attractor mechanism was first established for supersymmetric black holes \cite{Ferrara:1995ih,Strominger:1996kf,Ferrara:1996dd, Ferrara:1996um}, and later extended to non-supersymmetric extremal black holes \cite{Ferrara:1997tw,Gibbons:1997cc}.  At the two-derivative level,
the attractor equations can be obtained
 as extremization conditions for the effective potential of the physical scalar fields, known as the black hole potential \cite{Ferrara:1996dd,Ferrara:1997tw,Gibbons:1997cc,Goldstein:2005hq}. In the context of $N=2$ supergravity theories in four dimensions, this topic has been recently discussed in \cite{Kallosh:2005ax,Kallosh:2006bt,Bellucci:2006ew}.

A different way to understand the attractor mechanism is
the entropy function formalism \cite{Sen:2005wa,Sen:2005iz}.
In this approach, one defines an entropy function, whose extremization
determines the values of the scalar fields at
the horizon.  The entropy of the black hole is
then given by the value of the entropy function at the extremum.
For a specialization to four-dimensional $N=2$ supergravity
theories see \cite{Sahoo:2006rp}.

In the absence of higher curvature interactions,
the attractor equations can thus be derived both in the black hole
potential and in the entropy function approach.
In this paper, we contrast both methods in the setting of
$N=2$ supergravity in four dimensions.
We do this by explicitly
solving the attractor equations for the one-modulus prepotential
associated with the conifold of the mirror quintic in type IIB
\cite{Candelas:1990rm}.
We focus on extremal black holes carrying
two non-vanishing charges.  The advantage of the entropy function approach
is relative simplicity, allowing us to obtain exact solutions.  We find
two solutions to the attractor equations: a supersymmetric and a
non-supersymmetric one.  These two solutions and their entropies
are not related in a simple
way to one another, unlike in the class of extremal
type IIA (large volume) black
hole solutions carrying D0 and D4 charge.  There, the two entropies are mapped into another by reversing
the sign of the D0 charge \cite{Tripathy:2005qp,
Kallosh:2006bt}.

Usually one writes the black hole entropy as a function of the charges, but in the context of the entropic principle \cite{Ooguri:2005vr, Gukov:2005bg} one regards the entropy, through the attractor equations, as a function on the moduli space.
In the one-modulus case, an explicit expression exists for
the entropy of an extremal black hole
as a function of the scalar fields
\cite{Behrndt:1996jn,Bellucci:2006ew}.  For the conifold prepotential we find that, whereas
the entropy attains a local maximum at the conifold point for the BPS solution
\cite{Cardoso:2006nt} (see also \cite{Fiol:2006dv,Bellucci:2006ew}),
it possesses a local minimum there for the non-BPS solution.  Nonetheless,
the entropy of the non-BPS solution has a local maximum in the vicinity of the conifold point, and the point
corresponding to this local maximum represents a stable solution
to the attractor equations.  At this local maximum, the
entropy is larger than the entropy
of the BPS solution at the conifold point.

This paper is organized as follows.  In section 2 we review the
black hole potential formalism.
We solve the attractor equations for the conifold prepotential
in the leading-order approximation for the case of two non-vanishing
charges and we compute the entropy.
In section 3 we consider the entropy function formalism.
First we show that the entropy function is equivalent
to the black hole potential
and we give the associated attractor equations.
These results were
obtained in collaboration with Bernard de Wit and Swapna Mahapatra. Further
results and extensions, including the generalization
to $R^2$-interactions,
will appear
in a forthcoming publication \cite{cdwm}.
In the entropy function formalism the attractor equations for the
conifold prepotential and two non-zero charges take
a sufficiently simple form to allow manageable exact solutions with 
Mathematica,
which we display in subsection \ref{conent}.
In section 4 we
examine the stability of these solutions. And finally, in section 5,
we discuss the extrema of the entropy as a function of the modulus.

\section{Black hole potential approach}

In this section we review the black hole potential formalism and the
associated attractor equations
for four-dimensional static extremal black holes in $N=2$ supergravity theories
without higher-derivative interactions
\cite{Ferrara:1996dd,Ferrara:1997tw,Gibbons:1997cc}.
We then consider the conifold prepotential and we solve the attractor equations
for the case of two
non-vanishing charges.  We obtain two solutions: a supersymmetric and
a non-supersymmetric one, and we compute the associated entropies.
The two solutions differ substantially from one another.

\subsection{Review of black hole potential technology
\label{review}}

In four-dimensional $N=2$ supergravity coupled to $n$
Abelian vector multiplets \cite{deWit:1983rz,deWit:1984pk,deWit:1984px},
the central charge is given by
\begin{eqnarray}\label{central}
Z(X)=p^IF_I(X)-q_IX^I\;,\;\;\; I = 0, \dots, n \;.
\end{eqnarray}
Here $p^I$ and $q_I$ denote the set of magnetic and electric charges, which
we combine into the symplectic vector
\begin{eqnarray}\label{Q}
Q=\left(\begin{array}{c}p^I \\ q_I \end{array}\right).
\end{eqnarray}
The vector multiplet sector is determined by a holomorphic
prepotential $F(X)$, homogeneous of second degree,
with $F_I = \partial_I F(X)$, $F_{IJ} = \partial_I \partial_J F(X)$, etc.

Introducing the
symplectic vector
\begin{eqnarray}
V=\left(\begin{array}{c}X^I \\ F_I(X)\end{array}\right)
\end{eqnarray}
and the symplectic matrix
\begin{eqnarray}\label{Omega}
\Omega=\left(\begin{array}{cc} 0 & \mathbb{I} \\ -\mathbb{I} & 0\end{array}\right),
\end{eqnarray}
we can write (\ref{central}) as
\begin{eqnarray}
Z(X)=Q\T \, \Omega \, V \;.
\end{eqnarray}
The $X^I$ are related to the holomorphic sections $X^I(z)$ of special K\"ahler geometry by
\cite{deWit:1984px,Strominger:1990pd,Castellani:1990tp,Castellani:1990zd,
Candelas:1990pi,D'Auria:1990fj}
\begin{eqnarray}
X^I=\ee^{K(z,\bar z)/2} \, X^I(z)\spc ,
\end{eqnarray}
where $K(z,\bar z)$ denotes the K\"ahler potential,
\begin{eqnarray}
K(z,\bar z) = - \log \left( \im {\bar X}^I ({\bar z}) \, F_I (X(z)) - \im
X^I(z) \, {\bar F}_I ( {\bar X} ({\bar z}) ) \right) \;,
\end{eqnarray}
and
$z^A=X^A/X^0$
are the special coordinates of special K\"ahler geometry ($A=1,..., n$).
The number of physical scalar fields $z^A$ is one less than the number of 
$X^I$ because of the constraint
\begin{eqnarray}
\im \left( {\bar X}^I \, F_I (X) - X^I \, {\bar F}_I ({\bar X}) \right)=1
\label{constrai}
\end{eqnarray}
coming from the normalization of the Einstein term in the action.

Under K\"ahler transformations, the quantities
$X^I(z)$, $K$ and $Z(X)$ transform according to
\begin{eqnarray}\label{Ktrafo}
\quad X^I(z)\rightarrow \ee^{-f(z)}X^I(z)\spc ,\quad
K(z,\bar z)\rightarrow K(z,\bar z)+f(z)+\bar f(\bar z) \spc, \quad Z\rightarrow \ee^{-\frac{1}{2}\left[f(z)-\bar f(\bar z)\right]} \, Z \,.
\end{eqnarray}
The K\"ahler covariant derivative of $Z$ reads
\begin{eqnarray}
D_AZ(X) = \partial_A Z(X)
+\frac{1}{2}\left(\partial_A K\right)Z(X) \spc, \quad \partial_A=\frac{\partial}{\partial z^A} \spc \spc.
\end{eqnarray}
Observe that the $X^I$ are covariantly holomorphic, i.e.
${\bar D}_{\bar A} X^I =0$.

In the K\"ahler gauge $X^0(z)=1$, the K\"ahler potential is given by
\begin{equation}\label{kahlerpot}
\ee^{-K(z,\bar z)}=2\left(\cal F+\bar{\cal F}\right)-
\left(z^A - {\bar z}^A\right)
\left({\cal F}_A - \bar{{\cal F}}_{A}\right) \hspace{2mm} \spc \,,
\end{equation}
where $F(X) = - \im \,(X^0)^2 \, {\cal F}(z) $ and ${\cal F}_A = \partial
{\cal F}/ \partial z^A$.

In order to define the black hole potential, we introduce
the real matrix \cite{Ceresole:1995ca,Ferrara:1996dd}
\begin{eqnarray}\label{matrix}
M(\mathcal{N})=\left(\begin{array}{cc}\mathcal{I}+\mathcal{RI}^{-1}\mathcal{R} & -\mathcal{RI}^{-1} \\ -\mathcal{I}^{-1}\mathcal{R} & \mathcal{I}^{-1}\end{array}\right) \spc,
\end{eqnarray}
where
\begin{eqnarray}
\mathcal{R}=\Re\mathcal{N}  \spc, \quad
\mathcal{I}=\Im\mathcal{N}\spc,
\end{eqnarray}
and
\begin{eqnarray}
\mathcal{N}_{IJ}
=\bar F_{IJ}+2 \im \frac{\Im F_{IK}\Im F_{JL}X^KX^L}{\Im F_{MN}X^MX^N}\spc .
\end{eqnarray}
The black hole potential is then given by
\cite{Ferrara:1996dd,Ferrara:1997tw,Gibbons:1997cc}
\begin{eqnarray}
V_{\rm BH}= - \frac{1}{2} \, Q\T \,M({\cal N})\, Q \;
\end{eqnarray}
and describes the scalar-field dependent energy density of the vector fields.
It can be expressed in terms of the central charge $Z(X)$ and derivatives
thereof as follows \cite{Ceresole:1995ca,Ferrara:1996dd,Ferrara:1997tw}.
Using the special geometry identities (see \cite{Ceresole:1995ca})
\begin{eqnarray}
F_I&=&\mathcal{N}_{IJ}X^J \spc, \\
D_AF_I&=&\bar {\mathcal{N}}_{IJ} D_AX^J \spc, \\
-\frac{1}{2}\left(\Im\mathcal{N}\right)^{-1\,IJ}&=&\bar X^IX^J+g^{A\bar B}D_AX^I\bar D_{\bar B}\bar X^{J} \spc ,
\end{eqnarray}
we compute $ -\im Q+\Omega\, M({\cal N})\, Q$ and obtain
\cite{Bellucci:2006ew}
\begin{eqnarray}\label{identity}
-\im Q+\Omega\, M({\cal N}) \,
Q=2\left(Z(X)\bar V+g^{A\bar B}D_AV\bar D_{\bar B}\bar Z(\bar X)\right) \;.
\end{eqnarray}
Decomposing \eqref{identity} into imaginary and real part yields
\begin{eqnarray}
\label{ReId} -\im Q&=&Z(X)\bar V-\bar Z(\bar X)V+g^{A\bar B}\left(D_AV\bar D_{\bar B}\bar Z(\bar X)-D_AZ(X)\bar D_{\bar B}\bar V\right) \spc ,\\
\label{ImId}\Omega\, M({\cal N}) \,Q&=&Z(X)\bar V+\bar Z(\bar X)V+g^{A\bar B}\left(D_AV\bar D_{\bar B}\bar Z(\bar X)+D_AZ(X)\bar D_{\bar B}\bar V\right) \spc .
\end{eqnarray}
Contracting \eqref{ImId} with $Q\T \, \Omega$ results in
\begin{eqnarray}
\label{BHpot} V_{\rm BH}=
-\frac{1}{2}Q\T
M(\mathcal{N})Q=\abs {Z(X)}^2+g^{A\bar B}D_AZ(X)\bar D_{\bar B}\bar Z(\bar X) \spc,
\end{eqnarray}
where we used (\ref{central}).  This expresses the black hole potential $V_{\rm BH}$ in terms of $Z(X)$ and derivatives thereof.

Extrema of the black hole potential with respect to $z^A$ satisfy
\cite{Ferrara:1997tw}
\begin{eqnarray}
\label{extrcond} \partial_A V_{\rm BH}=0 \;\Leftrightarrow\;
2 \,{\bar Z}({\bar X}) \,
D_AZ(X)
+g^{B\bar C}\left(\mathcal{D}_AD_BZ(X)\right)\bar D_{\bar C}\bar Z(\bar X)=0 \spc ,
\end{eqnarray}
where $\mathcal{D}_AD_BZ=\left(D_A \,\delta_B^C-\Gamma^C_{AB}\right)D_CZ\spc$.
By virtue of the special geometry relation
\begin{eqnarray}
\mathcal{D}_AD_B V = \im \, C_{ABC} \, {\bar D}^C
{\bar V} \;\;\;,\;\;\;  {\bar D}^C = g^{C {\bar C}} {\bar D}_{\bar C} \;,
\end{eqnarray}
the double derivative in \eqref{extrcond} can be replaced by
\begin{eqnarray}
\label{ddz}
\mathcal{D}_AD_B Z(X) = \im \, C_{ABC} \, {\bar D}^C
{\bar Z}({\bar X}) \;,
\end{eqnarray}
where
\begin{eqnarray}
C_{ABC}=\ee^KF_{IJK}(X(z))
\,\frac{\partial X^I(z)}{\partial z^A}
\, \frac{\partial X^J(z)}{\partial z^B}\frac{\partial X^K(z)}{\partial z^C} \spc .
\end{eqnarray}

For later use we also note that \cite{Ceresole:1995ca}
\begin{eqnarray}\label{QMQidentity}
\frac{1}{2}Q\T\,M(F)\,Q=g^{A\bar B}D_AZ(X)\bar D_{\bar B}\bar Z(\bar X)-\abs {Z(X)}^2 \spc ,
\end{eqnarray}
where $M(F)$ is given by \eqref{matrix} with $\mathcal{N}_{IJ}$ replaced by $F_{IJ}$. This can be checked by using the identity \cite{deWit:1996ag}
\begin{eqnarray}
\left(N^{-1}\right)^{IJ}=g^{A\bar B}D_AX^I\bar D_{\bar B}\bar X^J-X^I\bar X^J \spc , \qquad N_{IJ}=\im\left(\bar F_{IJ}-F_{IJ}\right) \spc .
\end{eqnarray}

The black hole potential (\ref{BHpot}) is expressed in terms of $X^I$ and
$z^A$. It will be useful to express it in terms of rescaled
K\"ahler invariant variables
$Y^I$ \cite{Behrndt:1996jn},
\begin{eqnarray}\label{Pi}
\Pi(Y)=\bar Z(\bar X)V=\left(\begin{array}{c}Y^I \\ F_I(Y)\end{array}\right) \spc .
\end{eqnarray}
In terms of the $Y$-variables, equations \eqref{ReId} and \eqref{BHpot} become
\begin{eqnarray}
\label{ReId(Y)} -\im Q&=&\bar \Pi-\Pi+Z(Y)^{-1}g^{A\bar B}\left(\partial_A\Pi \,\bar \partial_{\bar B}\bar Z(\bar Y)-\partial_AZ(Y)\bar \partial_{\bar B}\bar \Pi\right) \spc , \\
\label{BHpotential(Y)}V_{\rm BH}&=&Z(Y)+Z(Y)^{-1}g^{A\bar B}\partial_AZ(Y)\bar \partial_{\bar B}\bar Z(\bar Y) \spc ,
\end{eqnarray}
respectively, where
\begin{eqnarray}\label{zy}
Z(Y)=Q\T \, \Omega \, \Pi = p^IF_I(Y)-q_IY^I \;.
\end{eqnarray}
Observe that $Z(Y)$ is real, i.e. $Z(Y) = |Z(X)|^2 =
{\bar Z}({\bar Y})$, and
that it may be written as \cite{Behrndt:1996jn}
\begin{eqnarray} \label{central3}
Z(Y) = |Z(X)|^2 \, \im \left(\bar X^I \, F_I  -X^I \bar F_I
\right)
= \im\left(\bar Y^I F_I(Y) -Y^I \bar F_I ({\bar Y})\right) =
|Y^0|^2 \,\ee^{-G(z,\bar z)} \;,
\end{eqnarray}
where we used the constraint \eqref{constrai}
in the first step,
and $Y^0 = \bar Z (\bar X) \, \ee^{K/2} \, X^0(z)$
in the last step, and where
\begin{eqnarray}
G(z,\bar z)=K(z,\bar z) +\log\abs{X^0(z)}^2 \spc.
\end{eqnarray}
In the K\"ahler gauge $X^0(z)=1$, $G=K$.

In the following, we will assume that $Z(Y) \neq 0$.
Inserting the extremization condition \eqref{extrcond},
\begin{eqnarray}
\partial_A Z(Y)= -\frac{1}{2Z(Y)}g^{B\bar C}
{\cal D}_A
D_BZ(Y)\,{\bar \partial}_{\bar C}\bar Z(\bar Y) \;,
\end{eqnarray}
into \eqref{ReId(Y)} yields the so-called attractor equations
\cite{Kallosh:2005ax},
\begin{eqnarray}\label{Attrequ}
Q=2\Im\left(\Pi(Y)+\frac{1}{2}
(Z(Y))^{-2}\,
g^{A\bar B}g^{\bar DE}\bar {\mathcal{D}}_{\bar B}\bar D_{\bar D}
\bar Z(\bar Y) \, \partial_A\Pi \, \partial_E
Z(Y)\right) \spc .
\end{eqnarray}
Using (\ref{ddz}), the
double derivative in \eqref{Attrequ} can, in the K\"ahler gauge
$X^0(z) =1$, be written as
\begin{eqnarray}\label{doubleD}
\mathcal{D}_AD_BZ(Y)=\im \,
\frac{\bar Z(\bar X)}{Z(X)}\,C_{ABC} \, g^{C {\bar C}} \, \partial_{\bar C}
{\bar Z}(\bar Y) =
\im \,\frac{Y^0}{{\bar Y}^0}\,C_{ABC} \, g^{C {\bar C}} \, \partial_{\bar C}
{\bar Z}(\bar Y)
\spc ,
\end{eqnarray}
where
\begin{eqnarray}\label{Y0}
Y^0= \bar Z(\bar X)X^0=\ee^{K/2}\bar Z(\bar X)
=\ee^K\left[p^I\bar F_I(\bar X(\bar z))-q_0-
q_A\bar z^A \right] \spc .
\end{eqnarray}

The entropy of an extremal black hole
is determined by the value of the black hole potential \eqref{BHpotential(Y)} at the extremum of the potential
\cite{Ferrara:1996dd,Ferrara:1997tw,Gibbons:1997cc},
\begin{eqnarray}\label{entropy1}
\Sc/\pi=V_{\rm BH}|_{\text{extr}}=\left.\left(Z(Y)+Z(Y)^{-1}g^{A\bar B}\partial_AZ(Y)\bar \partial_{\bar B}\bar Z(\bar Y)\right)\right|_{\text{extr}} \spc .
\end{eqnarray}
In the supersymmetric case, where $\partial_AZ=0$, the entropy reduces to
\begin{eqnarray}\label{entropy2}
\Sc_{\text{BPS}}/\pi=Z(Y)|_{\text{extr}} \spc .
\end{eqnarray}
On the other hand, in the non-supersymmetric case, and restricting to
prepotentials
${\cal F}(z)$ that only depend on one single modulus $z^1=z$,
it was shown in \cite{Bellucci:2006ew} that \eqref{entropy1} can be written as
\begin{eqnarray}\label{entropy5}
\Sc/\pi=Z(Y)
\left.\left(1+4\,\frac{g_{z\bar z}^{3}}{|C_{111}|^2}\right)\right|_{\text{extr}}
\spc .
\end{eqnarray}
Using \eqref{central3}, the entropy can be expressed as a function of
the modulus $z$,
\begin{eqnarray}
\label{entrozone}
\Sc/\pi=|Y^0|^2 \, \ee^{-G(z,\bar z)}\left(1+4\,\epsilon\,\frac{g_{z\bar z}^3}{|C_{111}|^2}\right) \spc ,
\end{eqnarray}
where $\epsilon=0,1$ for BPS and non-BPS black holes, respectively.

\subsection{Solving the attractor equations for the conifold prepotential \label{solutionsVBH}}

In this subsection,
we consider a specific one-modulus prepotential and solve
the attractor equations \eqref{Attrequ} following from the black hole
potential \eqref{BHpotential(Y)}, for two non-vanishing charges.
Then, using \eqref{entropy1},
we compute the entropy of the resulting black hole.
We refer to \cite{Tripathy:2005qp,Giryavets:2005nf,Kallosh:2006bt} for other
examples.

The prepotential we consider is the
conifold prepotential \cite{Candelas:1990rm}
\begin{eqnarray}\label{prepot}
F(Y)=-\im\left(Y^0\right)^2{\cal F} (T)=-\im\left(Y^0\right)^2\left[\frac{\beta}{2\pi}T^2\log T+a\right] \spc ,
\end{eqnarray}
where $T=-\im z = - \im Y^1/Y^0 \spc$,
$\beta$ is a real negative
constant and $a$ is a complex constant with $\Re a>0$.

For simplicity we consider extremal black holes with two non-vanishing
charges $q_0$ and $p^1$, so that
the charge vector $Q$ is given by
\begin{eqnarray}
\label{q2c}
Q=\left(\begin{array}{c}0 \\ p^1 \\ q_0 \\ 0 \end{array}\right).
\end{eqnarray}
In the following, we calculate all the quantities that
appear in \eqref{Attrequ} for the prepotential \eqref{prepot}.
We work in the K\"ahler gauge $X^0(z)=1$.
The resulting exact expressions are displayed
in subsection \ref{exact}. Since these expressions are complicated,
we approximate them in subsection \ref{approx} so as to be able to solve
\eqref{Attrequ}.

\subsubsection{Exact expressions\label{exact}}

Computing the derivative ${\cal F}_T=
\partial {\cal F}/\partial T = \beta T \left(2\log T+1\right)/(2 \pi) $
and inserting it into \eqref{kahlerpot} yields the K\"ahler potential
\begin{eqnarray}
K(T,\bar T)=-\log\left(4\Re a-\frac{\beta}{2\pi}\left(T+\bar T\right)^2-\frac{2\beta}{\pi}|T|^2\log|T|\right) \spc .
\end{eqnarray}
Computing
$Y^1=\im TY^0$, $F_0=\partial F/\partial Y^0=-2\im a Y^0 +\im
\beta T^2 \,Y^0/(2 \pi)$ and $F_1=\partial F/\partial Y^1=-\beta T \,Y^0
\log T/\pi
-\beta T  \,Y^0/(2 \pi)$, we obtain for the vector $\Pi (Y)$,
\begin{eqnarray}
\Pi(Y) =
\left(\begin{array}{c}Y^0 \\ Y^1 \\ F_0 \\ F_1 \end{array}\right)
=\left(\begin{array}{c}Y^0\\\im TY^0\\-2\im aY^0+\im\frac{\beta T^2}{2\pi}Y^0\\-\frac{\beta T}{\pi}Y^0\log T-\frac{\beta T}{2\pi}Y^0\end{array}\right) \spc .
\end{eqnarray}
The central charge \eqref{zy} takes the form
\begin{eqnarray}
\label{z2c}
Z(Y)=-\frac{\beta }{\pi}p^1Y^0 T \log T-\frac{\beta }{2\pi}p^1Y^0 T-q_0Y^0 \spc .
\end{eqnarray}
Using \eqref{Y0}, we obtain
\begin{eqnarray}
\partial_zZ(Y)&=&-\im\partial_TZ=\frac{\im\beta}{\pi}p^1Y^0\left(\log T+\frac{3}{2}\right)-\im\left(\partial_TK\right)Z \spc ,\\
\partial_z\Pi(Y)&=&-\im\partial_T\Pi=\left(\begin{array}{c}-\im\left(\partial_TK\right)Y^0\\Y^0\left(1+T\left(\partial_TK\right)\right)\\Y^0\left(-2a\left(\partial_TK\right)+\frac{\beta T}{\pi}+\frac{\beta T^2}{2\pi}\left(\partial_TK\right)\right)\\\frac{\im\beta}{\pi}Y^0\left(1+T\left(\partial_TK\right)\right)\log T+\frac{\im\beta}{2\pi}Y^0\left(3+T\left(\partial_TK\right)\right)\end{array}\right) \spc ,
\end{eqnarray}
where
\begin{eqnarray}
\partial_TY^0=\left(\partial_TK\right)Y^0=\frac{\frac{\beta Y^0}{\pi}\left(2\bar T+T+2\bar T\log|T|\right)}{4\Re a-\frac{\beta}{2\pi}\left(T+\bar T\right)^2-\frac{2\beta}{\pi}|T|^2\log|T|} \spc .
\end{eqnarray}
The metric $g_{T\bar T}=\partial_T\bar {\partial}_{\bar T}K$ is computed to be
\begin{eqnarray}
g_{T\bar T}=\frac{\frac{4\beta}{\pi}\Re a\left(3+2\log|T|\right)+\frac{\beta^2}{\pi^2}\left(\frac{1}{2}T^2+2|T|^2+\frac{1}{2}\bar T^2\right)+\frac{2\beta^2}{\pi^2}\left(T^2+\bar T^2\right)\log|T|}{\left(4\Re a-\frac{\beta}{2\pi}\left(T+\bar T\right)^2-\frac{2\beta}{\pi}|T|^2\log|T|\right)^2} \spc .
\end{eqnarray}
Using $\bar F(\bar X(\bar z)) = \im \,{\bar {\cal F}}
({\bar z})$, we have
\begin{eqnarray}
\bar C_{{\bar 1} {\bar 1} {\bar  1}}=\ee^K \,
\frac{\partial^3 \bar F(\bar X(\bar z))}{\partial \bar z^3}=\ee^K\frac{\beta}{\pi \bar T} \spc .
\end{eqnarray}
Inserting this into \eqref{doubleD} gives
\begin{multline}
\bar {\mathcal{D}}_{\bar z}\bar \partial_{\bar z}\bar Z(\bar Y)\\
=\frac{-\im\bar Y^0\left(4\Re a-\frac{\beta}{2\pi}\left(T+\bar T\right)^2-\frac{2\beta}{\pi}|T|^2\log|T|\right)\left(\frac{\im\beta}{\pi}p^1Y^0\left(\log T+\frac{3}{2}\right)-\im\left(\partial_TK\right)Z\right)}{Y^0\bar T\left(4\Re a\left(3+2\log|T|\right)+\frac{\beta}{\pi}\left(\frac{1}{2}T^2+2|T|^2+\frac{1}{2}\bar T^2\right)+\frac{2\beta}{\pi}\left(T^2+\bar T^2\right)\log|T|\right)} \spc .
\end{multline}

For the two-charge case \eqref{q2c},
the black hole potential \eqref{BHpotential(Y)} is invariant under the exchange
\begin{eqnarray}
\label{exch}
Y^0 \leftrightarrow \bar{Y}^0\spc,  \qquad  T \leftrightarrow \bar{T}
\spc.
\end{eqnarray}
This can be seen as follows.  Under the exchange \eqref{exch}, $Z(Y)$
given in \eqref{z2c} transforms into
\begin{eqnarray}
Z(Y) \rightarrow Z({\bar Y}) =
-\frac{\beta \bar T}{\pi}p^1\bar Y^0\log{\bar T}-\frac{\beta \bar T}{2\pi}p^1\bar Y^0-q_0\bar Y^0  = \bar Z(\bar Y)\spc .
\end{eqnarray}
On the other hand, since $Z(Y)$ is, by construction, a real quantity,
i.e. $Z(Y)= |Z(X)|^2 = {\bar Z} ({\bar Y}) $, it follows that
$Z(Y) = Z({\bar Y})$ under \eqref{exch}.
Similarly, $g^{T {\bar T}} \partial_T Z(Y) \partial_{\bar T} {\bar Z}
({\bar Y})$ is invariant under the exchange \eqref{exch}.
Thus, analogously to 
\cite{Sahoo:2006rp}, we will look for the class of solutions
to the
attractor equations  \eqref{Attrequ}
that are invariant under  \eqref{exch}, namely 
for solutions with real $Y^0$ and
real $T$.
To further ease the computations we also assume that $T\geq 0$ when solving
the attractor equations.

\subsubsection{Approximate solutions \label{approx}}

Now we approximate the expressions calculated in the last section by only keeping the leading terms in the limit $T\rightarrow 0$, i.e.\ we consider $T$ in the vicinity of the conifold point. We obtain
\begin{eqnarray}
\label{metric2}g_{z\bar z}&\approx&\frac{\beta}{2\pi \Re a}\log|T| \spc ,\\
\partial_TY^0&\approx&\frac{\beta Y^0\bar T\log |T|}{2\Re a} \spc ,\\
Z(Y)&\approx&-\frac{\beta p^1Y^0}{\pi}T\log T-q_0Y^0 \spc ,\\
\partial_zZ(Y)&\approx&\frac{\im\beta p^1Y^0}{\pi}\log T+\frac{\im\beta q_0Y^0}{2\Re a}\bar T\log |T|  \spc ,\\
\bar {\mathcal{D}}_{\bar z}\bar \partial_{\bar z}\bar Z(\bar Y)&\approx&\frac{\bar Y^0\beta\left(\frac{p^1}{\pi}\log T+\frac{q_0}{2\Re a}\bar T\log |T|\right)}{2\bar T\log|T|} \spc ,\\
\label{DPi2}\partial_z\Pi(Y)&\approx&\left(\begin{array}{c}-\frac{\im\beta Y^0\bar T\log |T|}{2\Re a}\\Y^0\\-\frac{a\beta Y^0}{\pi\Re a}\bar T\log |T|\\\frac{\im\beta Y^0}{\pi}\log T\end{array}\right) \spc .
\end{eqnarray}
Setting $Y^0 = {\bar Y}^0$ and $T = {\bar T}$ as mentioned above,
we then obtain for the attractor equations \eqref{Attrequ},
\begin{multline}
\left(\begin{array}{c}0 \\ p^1 \\ q_0 \\ 0 \end{array}\right)=2\Im\left[\left(\begin{array}{c}Y^0\\ \im TY^0\\-2\im aY^0+\im\frac{\beta T^2}{2\pi}Y^0\\-\frac{\beta T}{\pi}Y^0\log T-\frac{\beta T}{2\pi}Y^0\end{array}\right)\right]\\
+2\Im\left[\frac{\im\pi^2\left(2\Re {a} \, p^1+q_0T\right)^2}{4T\log T\left(\beta p^1T\log T+\pi q_0\right)^2}\left(\begin{array}{c}-\frac{\im\beta Y^0 T\log T}{2\Re a}\\Y^0\\-\frac{a\beta Y^0}{\pi\Re a} T\log T\\\frac{\im\beta Y^0}{\pi}\log T\end{array}\right)\right] \spc .
\end{multline}
Taking the imaginary part results in the following two attractor equations
involving $p^1$ and $q_0$,
\begin{subequations}
\begin{align}
p^1 &= 2TY^0+Y^0\frac{\pi^2\left(2\Re {a} \, p^1+q_0T\right)^2}{2T\log T\left(\beta p^1T\log T+\pi q_0\right)^2} \spc ,\\
q_0 &= -4\Re a Y^0+\frac{\beta T^2}{\pi}Y^0-\frac{a\beta\pi Y^0\left(2\Re{a} \, p^1+q_0T\right)^2}{2\Re{a}\left(\beta p^1T\log T+\pi q_0\right)^2} \spc .
\end{align}
\end{subequations}

These equations can be approximately solved to leading order in $T$ in the limit $T \rightarrow 0$.  We find the following two approximate solutions, one
of which is supersymmetric.  The supersymmetric one ($\partial_z Z(Y) =0$) is given by \cite{Cardoso:2006nt}
\begin{subequations}
\label{pBPS}
\begin{align}
p^1
&\approx 2Y^0T \spc,\\
q_{0
}&\approx -4\hspace{1mm}{\rm Re}\hspace{1mm}a\hspace{1mm}Y^0 \spc ,
\end{align}
\end{subequations}
whereas the non-supersymmetric solution  ($\partial_z Z(Y) \neq 0$) reads
\begin{subequations}
\label{qnonBPS}
\begin{align}
p^1
&\approx 8Y^0T\log T \spc ,\\
q_{0
}&\approx -4\hspace{1mm}{\rm Re}\hspace{1mm}a\hspace{1mm}Y^0 \spc .
\end{align}
\end{subequations}
Solving for the modulus $T$ yields
\begin{subequations} \label{apprtsu}
\begin{align}
Y^0
&\approx -\frac{q_0}{4\hspace{1mm}{\rm Re}\hspace{1mm}a\hspace{1mm}} \spc ,\\
T
&\approx -2\hspace{1mm}{\rm Re}\hspace{1mm}a\hspace{1mm}\frac{p^1}{q_0} \spc ,
\end{align}
\end{subequations}
in the supersymmetric case, and
\begin{subequations}\label{tlt}
\begin{align}
Y^0
&\approx-\frac{q_0}{4\hspace{1mm}{\rm Re}\hspace{1mm}a\hspace{1mm}} \spc ,\\
T\log T
&\approx-\frac{1}{2}\hspace{1mm}{\rm Re}\hspace{1mm}a\hspace{1mm}\frac{p^1}{q_0} \spc .
\end{align}
\end{subequations}
in the non-supersymmetric case.

The conditions $T\ll 1$ and $T\geq0$ constrain the choice of the charges. In the supersymmetric case, we have to choose $p^1$ and $q_0$ such that $p^1/q_0<0$ and $\abs{p^1}\ll\abs{q_0}$, whereas in the non-supersymmetric case the charges have to satisfy
$p^1/q_0>0$ and $\abs{p^1}\ll\abs{q_0}$.

Next, we calculate the entropy
in the limit $T \rightarrow 0$.  Inserting \eqref{pBPS} into
\eqref{entropy2} yields
\begin{eqnarray} \label{entroapprs}
\Sc/\pi
\approx\left(Y^0\right)^2\left(4\hspace{1mm}{\rm
    Re}\hspace{1mm}a\hspace{1mm}-\frac{2\beta}{\pi}T^2 \,\log T \right) \spc
\end{eqnarray}
in the supersymmetric case \cite{Cardoso:2006nt}, whereas inserting
\eqref{qnonBPS} into \eqref{entropy1} yields
\begin{eqnarray}\label{entropy4}
\Sc/\pi
\approx \left(Y^0\right)^2\left(4\hspace{1mm}{\rm Re}\hspace{1mm}a\hspace{1mm}+\frac{32\beta}{\pi}T^2\left(\log T\right)^3\right) \spc
\end{eqnarray}
in the non-supersymmetric case.  Both are
in accordance with the expression obtained from
\eqref{entrozone}.

We observe that the solutions \eqref{apprtsu} and \eqref{tlt} (and their
associated entropies \eqref{entroapprs} and \eqref{entropy4}) are not related
in a simple way to one another, in contrast to what happens in\
the case of cubic prepotentials \cite{Tripathy:2005qp,Kallosh:2006bt}.
For the cubic prepotential $F(Y) = - (Y^1)^3/Y^0$, the supersymmetric
solution to the attractor equations \eqref{Attrequ} is given by
$Y^0 = p^1/(2 T)\,, \,T = \sqrt{|q_0/p^1|}$ with $q_0/p^1 <0$,
whereas the non-supersymmetric solution is given by $Y^0 = p^1/(4 T) \,,\,
T=\sqrt{q_0/p^1}$ with $q_0/p^1 >0$.  The values of the $T$-modulus are mapped
into one another under $q_0 \rightarrow - q_0$.  The associated entropies,
${\cal S} = 2 \pi \sqrt{|q_0 \, (p^1)^3|}$ and
${\cal S} = 2 \pi \sqrt{q_0 \, (p^1)^3}$, respectively, are also mapped
into one another under this transformation.  For the
case of the conifold prepotential,
there is no such simple transformation relating the two solutions given above.

\section{
Entropy function approach}

In the following, we show that the entropy function \cite{Sen:2005wa, Sen:2005iz,Sahoo:2006rp}
evaluated for an $N=2$ supergravity Lagrangian without $R^2$-interactions
is equivalent to the black hole potential,
and we give the associated attractor equations (see also \cite{Alishahiha:2006jd}).
These results were
obtained in collaboration with Bernard de Wit and Swapna Mahapatra. Further
results and extensions, including the generalization
to $R^2$-interactions,
will appear
in a forthcoming publication \cite{cdwm}.

The advantage of the entropy function approach
over the black hole potential method is relative simplicity, as the formulation does not involve covariant derivatives nor mixing of $Y^I$ (or $X^I$) and $z^A$ variables (the natural variables for the central charge are $X^I$, while the differentiation in the black hole potential approach is with respect to $z^A$). What is more, the entropy function readily lends itself to the inclusion of higher-order corrections.

\subsection{The entropy function and the black hole potential}

Instead of using the black hole potential as a starting point, one can equivalently derive the attractor equations  from the entropy function
\cite{Sen:2005wa, Sen:2005iz},
defined as the Legendre transform of the Lagrangian density $\sqrt{-g}\,\mathcal{L}$ integrated at the horizon 
over the angular coordinates, with respect to the electric fields,
\cite{Sahoo:2006rp}
\begin{equation}
\Ec = 2\pi\left(-\frac{1}{2} q_I e^I - \int\de\theta\,\de\phi\,\sqrt{-g}\,\mathcal{L}\right),
\end{equation}
where $g$ denotes the determinant of the near-horizon metric on $AdS_2\times S^2$,
\begin{equation}
\de s^2 = v_1 (-r^2 \de t^2 + \de r^2/r^2) + v_2 (\de\theta^2 + \sin^2\!\theta\,\de\phi^2)\spc,
\end{equation}
and where
$e^I$ are electric fields and $q_I$ are the electric charges carried by the
extremal black hole. $\Ec$ has the same property as $\VBH$, namely its  extremization determines the values at the horizon of the parameters describing a black hole, while the extremum itself yields the entropy.  In fact, $\Ec$ and $\VBH$
are completely equivalent.

The entropy function evaluated for an $N=2$ supergravity Lagrangian, here considered without $R^2$ corrections,
can be rewritten, after the electric fields
$e^I$ have been eliminated through their equations of motion, entirely in terms of the $Y^I$ variables and the charges
(see eq.~(3.11) in \cite{Sahoo:2006rp}),
\begin{equation}
\Ec(Y,\bar{Y},p,q)/\pi = Q\T m(\tau)\,Q - 2\im\,Q\T m(\tau)\,\bar{\Pi} + 2\im\,\Pi\T m(\tau)\,Q - 2\im\,\bar\Pi\T\Omega\,\Pi\spc,
\end{equation}
where $Q$, $\Pi$ and $\Omega$ have been defined in \eqref{Q}, \eqref{Pi} and \eqref{Omega}, respectively,
\begin{equation}
m(\tau) =
\begin{pmatrix}
  \bar{\tau} N^{-1} \tau  &  -\bar{\tau} N^{-1} \\
  -N^{-1} \tau &   N^{-1}
\end{pmatrix},
\qquad
\tau_{IJ} \equiv F_{IJ}\spc, \qquad N = \im(\bar{\tau} - \tau) \spc,
\end{equation}
and the relationship between the $Y$-variables used here
and those of \cite{Sahoo:2006rp} is
\begin{equation}
\begin{pmatrix} Y^I \\ F_I \end{pmatrix} \leftrightarrow
\frac{v \, \bar{w}}{4} \begin{pmatrix} x^I \\ F_I \end{pmatrix}_\text{\cite{Sahoo:2006rp}} \qquad
(v_1 = v_2 = v)\spc.
\end{equation}
Due to the homogeneity of the prepotential, $\tau$ is homogeneous of degree $0$, so it is not subject to rescaling. Observe that $m(\tau)$ is Hermitian ($m\T = \bar{m}$), because $\tau$ is symmetric and $N$ is real (so that  $N^{-1}$ is both real and symmetric).

Thanks to these properties, as well as $m - \bar{m} = \im\Omega$, and the fact that a transpose of a scalar is equal to itself,
$\Ec$ can be recast into the form
\begin{equation}
\label{eq:EcIntermediate}
\begin{split}
\Ec(Y,\bar{Y},p,q)/\pi &= \frac{1}{2} Q\T M(\tau)\,Q + \im Q\T M(\tau)\,(\Pi-\bar\Pi) + Q\T\Omega\,(\Pi+\bar\Pi)
- 2\im\,\bar\Pi\T\Omega\,\Pi\\
&= \frac{1}{2} \left(Q + \im (\Pi-\bar\Pi)\right)\T M(\tau) \left(Q + \im (\Pi-\bar\Pi)\right)
+ \frac{1}{2}(\Pi-\bar\Pi)\T M(\tau)\,(\Pi-\bar\Pi)\\
&\quad + Q\T\Omega\,(\Pi+\bar\Pi) -2\im\,\bar\Pi\T\Omega\,\Pi \spc,
\end{split}
\raisetag{4ex}
\end{equation}
where
$M(\tau) = m(\tau) + \bar{m}(\tau)$ is the same symmetric matrix as in \eqref{matrix}, but evaluated for $F_{IJ}$ instead of $\mathcal{N}_{IJ}$,
\begin{gather}
M(\tau) =
\begin{pmatrix}
  \I + \R \I^{-1} \R  &  -\R \I^{-1} \\
  -\I^{-1} \R  &  \I^{-1}
\end{pmatrix} =
\begin{pmatrix}
  \bar{\tau} N^{-1} \tau + \tau N^{-1} \bar{\tau}  &  -(\tau + \bar{\tau}) N^{-1} \\
  -N^{-1} (\tau + \bar{\tau})  &  2 N^{-1}
\end{pmatrix}\spc, \label{eq:Mtau}\\
\R = \Re\tau\spc, \qquad \I = \Im\tau\spc.
\end{gather}

By direct expansion, exploiting the homogeneity relation
\begin{equation}
\label{eq:homoFIJ}
Y^I F_{IJ} = F_J\spc,
\end{equation}
the symmetry in indices and the definition of $N$ we have
\begin{equation}
\label{eq:PiMPi}
\frac{1}{2}(\Pi-\bar\Pi)\T M(\tau) \,(\Pi-\bar\Pi) = \im\bar\Pi\T\Omega\,\Pi
\spc.
\end{equation}
As a result the entropy function can be represented as a sum of two entities
\begin{gather}
\label{eq:Ec}
\Ec(Y,\bar{Y},p,q)/\pi= \Sigma(Y,\bar{Y},p,q) + \frac{1}{2} \left(Q + \im (\Pi-\bar\Pi)\right)\T M(\tau) \left(Q + \im (\Pi-\bar\Pi)\right),
\end{gather}
where \cite{Behrndt:1996jn}
\begin{eqnarray}
\label{sig}
\Sigma(Y,\bar{Y},p,q) &=& -\im\bar\Pi\T\Omega\,\Pi + Q\T\Omega\,(\Pi+\bar\Pi)
\nonumber\\
&=& -\im (\bar{Y}^I F_I - Y^I \bar{F}_I) + p^I (F_I + \bar{F}_I) - q_I (Y^I + \bar{Y}^I)\spc.
\end{eqnarray}
We will now identify \eqref{eq:Ec}
with the two parts of the black hole potential
\eqref{BHpotential(Y)}.

Substitution of \eqref{central3} into $\Sigma$ implies that
\begin{equation}
\Sigma(Y,\bar{Y},p,q) = \im (\bar{Y}^I F_I - Y^I \bar{F}_I) = Z(Y)
\spc,
\end{equation}
in which we recognize the first term in $\VBH$.

The product in \eqref{eq:Ec} decomposes into (cf.~\eqref{eq:EcIntermediate})
\begin{equation}
\label{eq:4Terms}
\frac{1}{2}\left[Q\T M(\tau)\,Q + \im Q\T M(\tau)\,(\Pi-\bar\Pi)
+ \im (\Pi-\bar\Pi)\T M(\tau)\,Q - (\Pi-\bar\Pi)\T M(\tau)
\,(\Pi-\bar\Pi) \right].
\end{equation}
The first term above becomes, by virtue of \eqref{QMQidentity},
\begin{equation}
\frac{1}{2} Q\T M(\tau)\,Q
= Z(Y)^{-1} g^{A\bar{B}} \partial_A Z(Y) \bar{\partial}_{\bar{B}}\bar{Z}(Y) - Z(Y)\spc.
\end{equation}
To be precise, the quoted identity concerns $M(\tau(X))$ and not $M(\tau(Y))$ as here, but as we have indicated, $\tau$ is homogeneous of degree $0$ and $\tau(X) = \tau(Y)$.

Recall from \eqref{eq:EcIntermediate} that as a consequence of $M = M\T$, the second and third term in \eqref{eq:4Terms} are equal to one another. Applying the same techniques as in the derivation of \eqref{eq:PiMPi} we obtain
\begin{equation}
\frac{\im}{2} (\Pi-\bar\Pi)\T M(\tau)\,Q = \frac{1}{2} \left(p^I  (F_I+\bar{F}_I)  - q_I (Y^I + \bar{Y}^I)  \right) = \frac{1}{2}\left(Z(Y) + \bar{Z}(Y)\right) = Z(Y)\spc,
\end{equation}
where we used that $Z(Y)$ is real.

Finally, collecting the partial results confirms that
\begin{equation}
\Ec/\pi = Z(Y) + Z(Y)^{-1} g^{A\bar{B}} \partial_A Z(Y) {\bar \partial}_{\bar{B}}\bar{Z}({\bar Y}) = \VBH\spc.
\end{equation}

\subsection{Attractor equations }

The attractor equations can be derived by demanding that $\Ec$, given in
\eqref{eq:Ec} and \eqref{sig},
 be stationary with respect to independent variations in $Y^I$ and $\bar{Y}^I$.
They read
\begin{equation}
\label{eq:NSAE}
2 v_I + \bar{v}_J N^{JK} F_{KIL}\pt^L - \im \bar{v}_J N^{JK} F_{KIL} N^{LM} v_M = 0\spc,
\qquad v_I = \qt_I - F_{IJ} \pt^J \;,
\end{equation}
(together with their complex conjugates),
where
\begin{equation} \label{quant}
\qt_I = q_I - 2\Im F_I\spc, \qquad \pt^I = p^I - 2\Im Y^I\spc, \qquad N^{IJ} \equiv (N^{-1})^{IJ}.
\end{equation}
Note that the quantities \eqref{quant} are real.
BPS attractors satisfy \cite{Behrndt:1996jn}
\begin{equation}
\label{eq:SAE}
\pt^I = 0\spc, \qquad \qt_I = 0\spc.
\end{equation}

\subsection{Solutions for the conifold prepotential and two charges \label{conent}}

In the entropy function formalism the attractor equations for the conifold prepotential \eqref{prepot} and two non-zero charges
$q_0$ and $p^1$ take a sufficiently simple form to allow a manageable exact solution. The system of equations \eqref{eq:NSAE} reduces (under the same simplifying assumptions as in subsection
\ref{approx}, namely
$Y^0 = {\bar Y}^0, T = {\bar T}$
and $T\geq 0$) to two independent simultaneous equations,
\begin{subequations}
\begin{align}
8 \beta Y^0 T^2 - 8 \beta p^1 T + 8 \pi q_0 + 32 \pi \Re a\, Y^0
+\frac{\beta (T (\beta p^1 T - 2 \pi q_0) - 4 \pi p^1 \Re a)^2}{Y^0 \left(\beta T^2 + 4 \pi \Re a \right)^2}
&= 0\spc,\\
4 p^1 (2 \log T + 3) - 8 Y^0 T (2 \log T + 3)
-\frac{(T (\beta p^1 T - 2 \pi q_0) - 4 \pi p^1 \Re a)^2}{Y^0 T \left(\beta T^2 + 4 \pi \Re a\right)^2} &= 0\spc,
\end{align}
\end{subequations}
which, as we have already seen, possess two (pairs of) solutions: one preserving half of supersymmetries and one supersymmetry-breaking.

The BPS solution
\begin{subequations}
\label{eq:BPSsolution}
\begin{align}
p^1 &= 2 Y^0 T \spc,\\
q_0 &= - 4 \Re a\, Y^0 + \frac{\beta}{\pi} Y^0 T^2
\end{align}
\end{subequations}
can be directly compared with the approximate solution \eqref{pBPS}.

For comparison with \eqref{qnonBPS} the exact non-supersymmetric solution
\begin{subequations}
\label{eq:nonBPSsolution}
\begin{align}
p^1 &= \frac{Y^0 T\left(\beta^2 T^4 (2\log T (\log T + 3) + 5) + 8\pi\Re a
  \left(\beta T^2(\log T + 2) + \pi \Re a (4 \log T + 7) \right)\right)}
  {\left(\beta T^2 (\log T + 1) - 2\pi\Re a \right)^2} \spc,\raisetag{6ex}\\
\begin{split}
q_0 &= -\frac{Y^0}{2 \pi \left(\beta T^2 (\log T + 1) - 2 \pi \Re a\right)^2}\\
&\quad\Bigl[\beta^3 T^6 \left(2 \log T (\log T + 2) + 1\right)
+ 8 \pi \Re a\,\Bigl(\beta^2 \, T^4 \left(\log T (5 \log T + 11) + 5\right)\\
&\quad+\pi \Re a \left(\beta T^2 \left(4 \log T (2 \log T + 3) + 1\right)
+ 4 \pi \Re a \right)\Bigr)\Bigr]\spc,
\end{split}
\end{align}
\end{subequations}
needs to be expanded for small $T$,
\begin{subequations}
\begin{align}
p^1 &= 8 Y^0 T \log T + 14 Y^0 T  + \bigO(T^3) \spc,\\
q_0 &= -4\Re a\, Y^0 - \frac{\beta}{\pi}Y^0 \,T^2 \,
\left(8 (\log T)^2 + 16\log T +5\right) + \bigO(T^3)
\end{align}
\end{subequations}
and turns out not to be related in a simple way to the BPS solution, as
we already mentioned when discussing the approximate result \eqref{qnonBPS}
(see the end of subsection \ref{approx}).

\section{Stability of solutions}

To verify whether a solution to the attractor equations is stable, viz.~indeed represents an attractor, one needs to check if it furnishes the black hole potential, regarded as a function of the moduli for a given set of charges, with a minimum. In practice it might be again more feasible to avail oneself of the entropy function \eqref{eq:Ec}, where now
$Y^1$ has been replaced by $\im T Y^0$, and $Y^0$ is (in the K\"ahler gauge $X^0(z)=1$)  expressed in terms of $T$ as
$Y^0 = \ee^{K(T,\bar T)} \left(-\frac{\beta}{\pi} p^1 \bar T(\frac{1}{2} + \log\bar T) - q_0 \right)$.

The quality of critical points of a real-valued function can be determined with the aid of its Hessian matrix (unless the second derivatives vanish). If, as here, the function has complex arguments $z^A$, by ``Hessian matrix'' we mean the Hessian computed with respect to real variables ($x^A = \frac{1}{2}(z^A + \bar{z}^A)$ and $y^A = \frac{1}{2\im}(z^A - \bar{z}^A)$), which may be expressed by the matrix of complex derivatives (using $\pd_x = \pd + \bar{\pd}$ and $\pd_y = \im(\pd - \bar{\pd})$) through the following block-matrix equation
\begin{equation}
\begin{pmatrix}
\frac{\pd^2 f}{\pd x^2} & \frac{\pd^2 f}{\pd x\pd y} \\
\frac{\pd^2 f}{\pd y \pd x} & \frac{\pd^2 f}{\pd y^2}
\end{pmatrix}
=
\begin{pmatrix}
\mathbb{I} & \im \\
\mathbb{I} & -\im
\end{pmatrix}^{\!\!\mathrm{T}}
\begin{pmatrix}
\frac{\pd^2 f}{\pd z^2} & \frac{\pd^2 f}{\pd z\pd\bar{z}} \\
\frac{\pd^2 f}{\pd z\pd\bar{z}} & \frac{\pd^2 f}{\pd \bar{z}^2}
\end{pmatrix}
\begin{pmatrix}
\mathbb{I} & \im \\
\mathbb{I} & -\im
\end{pmatrix}.
\end{equation}
(See also \cite{Bellucci:2006ew}). Whenever the Hessian is positive definite at a stationary point, this point must be a minimum.

\begin{figure}[t]
\centering
\includegraphics[bb=91 3 322 146]{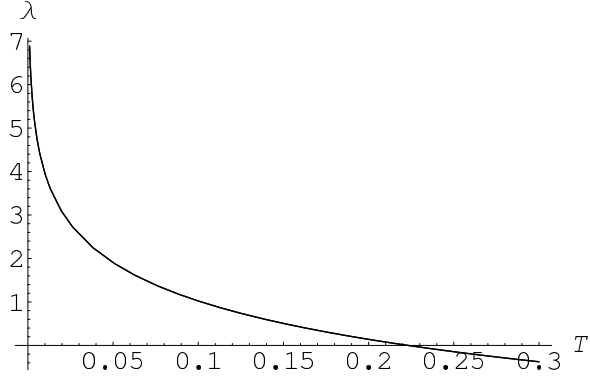}
\caption{Eigenvalue(s) $\lambda$ of the Hessian of the black hole potential
as a function of $T$ for $Y^0 =1, \beta=-1/2$, $a=1$ in the BPS case.}
\label{fig:susyEV}
\end{figure}

The Hessian matrix of $\VBH(z,\bar z,p,q)$ with the BPS solution \eqref{eq:BPSsolution} substituted after differentiation has one double eigenvalue, positive for sufficiently small $T = -\im z$ (Fig.~\ref{fig:susyEV}). This means that the BPS solution is stable, in accordance with the universal statement \cite{Ferrara:1997tw} that for supersymmetric solutions (the relevant part of) the Hessian is proportional to the K\"ahler metric, rendering all BPS solutions attractors, as long as the metric remains positive definite.

\begin{figure}[ht]
\centering
\includegraphics[bb=91 3 322 146]{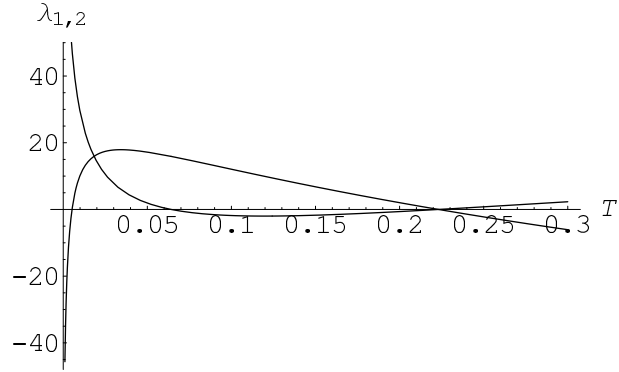}
\caption{Eigenvalues $\lambda_{1,2}$ of the Hessian of the black hole potential as
functions of $T$ for $Y^0=1$, $\beta=-1/2$, $a=1$ in the non-BPS case.}
\label{fig:nonsusyEV}
\end{figure}

For the non-BPS solution \eqref{eq:nonBPSsolution} the Hessian of the black hole potential has two distinct eigenvalues, which exhibit complicated behavior as $T$ varies (Fig.~\ref{fig:nonsusyEV}). For very small $T$ the eigenvalues have opposite signs, indicating a saddle point of the potential (the solution is unstable), but in the range approximately $T\in [0.005,0.06]$ the eigenvalues are both positive, so the solution becomes an attractor (provided that the prepotential in this region can be still reliably described by \eqref{prepot}).

\section{Extrema of the entropy in the moduli space}

In the context of the entropic principle \cite{Ooguri:2005vr, Gukov:2005bg} one is interested in the black hole entropy as a function on the moduli space, rather than, as otherwise common, a function of the charges. (We ignore at this point the question whether a solution to the attractor equations with integral charges can be found for an arbitrary point in the moduli space.)

\begin{figure}[ht]
\includegraphics[width=.48\linewidth,bb=91 3 322 146]{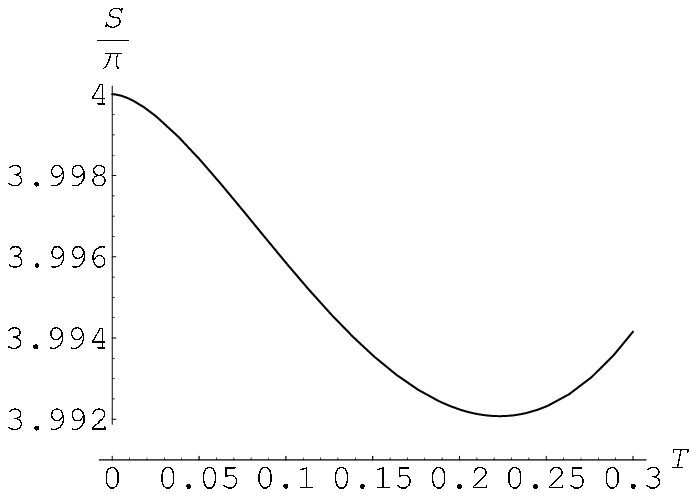}
\hspace{\stretch{1}}
\includegraphics[width=.48\linewidth,bb=91 21 322 172]{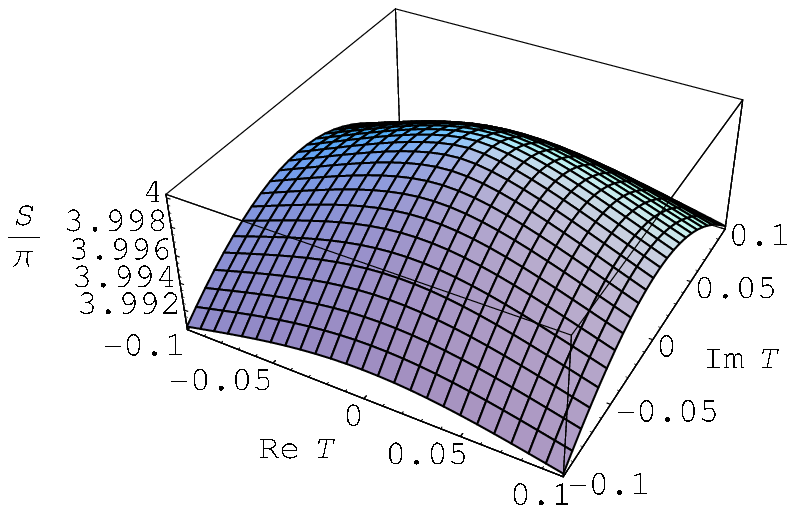}
\caption{$\Sc/\pi$ as a function of $T$ for $Y^0=1$, $\beta=-1/2$, $a=1$ in the BPS case. The left graph is a cross section along the positive $T$ semi-axis through the surface in the right graph.}
\label{fig:susyS}
\end{figure}

Inserting the BPS solution \eqref{eq:BPSsolution} into \eqref{eq:Ec} yields the entropy
\begin{equation}
\label{eq:SBPSexpl}
\Sc = 2 (Y^0)^2 \left(2 \pi \Re a - \beta T^2 (\log T + 1)\right),
\end{equation}
in agreement with \eqref{entrozone},
\begin{equation}
\label{eq:SBPSgen}
\Sc_{\text{BPS}} = \pi \abs{Y^0}^2 \ee^{-G(z,\bar z)}
= 2 \abs{Y^0}^2 \left(2\pi\Re a - \beta\abs{T}^2\log\abs{T} - \beta (\Re T)^2\right).
\end{equation}
The entropy
$\Sc$, regarded as a function of $z$ (or $T$) for constant $Y^0$, has a local maximum at the conifold point $T=0$ \cite{Cardoso:2006nt}, as shown in Fig.~\ref{fig:susyS}. The left graph corresponds to our explicit solution \eqref{eq:SBPSexpl} for two charges (constrained to the positive $T$ semi-axis), while the right represents the general formula \eqref{eq:SBPSgen}, without restrictions on the charges.

\begin{figure}[h]
\includegraphics[width=.48\linewidth,bb=91 3 322 146]{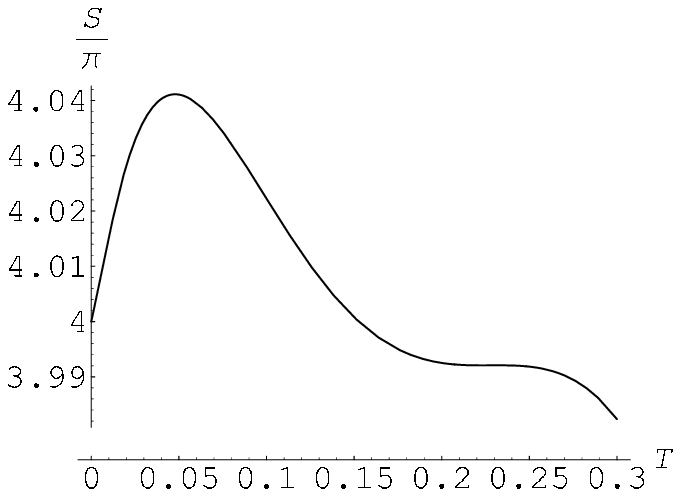}
\hspace{\stretch{1}}
\includegraphics[width=.48\linewidth,bb=91 21 322 172]{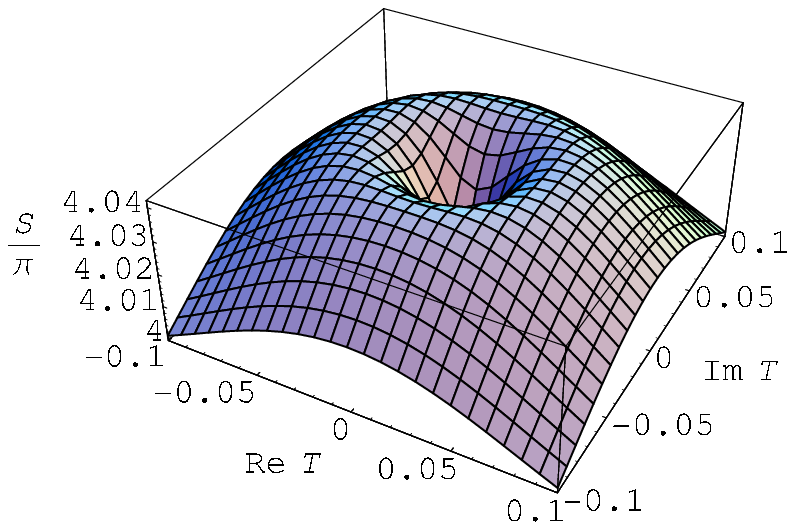}
\caption{$\Sc/\pi$ as a function of $T$ for $Y^0=1$, $\beta=-1/2$, $a=1$ in the
non-BPS case, similarly to Fig.~\ref{fig:susyS}.}
\label{fig:nonsusyS}
\end{figure}

Inserting the non-BPS solution \eqref{eq:nonBPSsolution} into \eqref{eq:Ec} yields the entropy
\begin{equation}
\begin{split}
\frac{\Sc}{\pi (Y^0)^2} & = \frac{1}{4 \pi  \left(\beta T^2 (\log T + 1) - 2 \pi \Re a\right)^4}\\
&\quad\Biggl[\frac{1}{\beta T^2 + 4 \pi \Re a}%
\Biggl(\beta T^2 (2 \log T + 3)^3 \left(\beta T^2 + 4\pi\Re a\right)^5\\
   &\qquad-{}\Bigr(\beta^2 T^4 \left(-9 \beta T^2 + 4 \left(a \pi - \beta T^2 \right) (\log T)^2
   + 4 \left(2 a \pi -3 \beta T^2 \right) \log T + 4 a \pi \right) \\
   &\qquad-8\pi \beta \left. \Re a\,T^2\left(5 \beta T^2 (\log T)^2 + 2 \left(7 \beta T^2 + a \pi \right) \log T
   + 2 \left(5 \beta T^2 + a \pi \right)\right) \right.\\
   &\qquad - 32 \pi ^3 (\Re a)^3 + 16 \pi^2 (\Re a)^2 \left(-4 \beta T^2 (\log T)^2 - 7 \beta T^2
   - 10 \beta T^2 \log T + a \pi \right)\\
   &\qquad+ 4 \pi \bar a \left(\beta T^2 (\log T + 1) - 2 \pi \Re a\right)^2\Bigr)^2\Biggr)
   - 8 \left(\beta T^2 (\log T + 1) - 2 \pi \Re a\right)^5 \Biggr]  \\
 & = 4\Re a + \frac{\beta}{\pi} T^2 \left(32(\log T)^3 + 144(\log T)^2 + 214\log T + 106\right) + \bigO(T^3) \spc,
\raisetag{4.4ex}
\end{split}
\end{equation}
again in agreement with \eqref{entrozone},
\begin{equation}
\begin{split}
\Sc_{\text{non-BPS}} &= \pi \abs{Y^0}^2\ee^{-G(z,\bar z)}\left(1+4\frac{g_{z\bar z}^3}{\abs{C_{111}}^2}\right)\\
&= 2\abs{Y^0}^2 \left(2\pi\Re a - \beta\abs{T}^2\log\abs{T} - \beta(\Re T)^2\right)\\
&\hskip -5mm
\left(1 + \beta \abs{T}^2\frac{\left(\beta \abs{T}^2(1 - 2 \log\abs{T})
  + 2 \beta (\Re T)^2\,(1 + 2 \log\abs{T}) + 4\pi\Re a\,(3+ 2\log\abs{T})\right)^3}
  {4\left(2\pi\Re a - \beta\abs{T}^2\log\abs{T} - \beta(\Re T)^2\right)^4}\right).
\end{split}
\raisetag{13ex}
\end{equation}
In contrast to the BPS case, $\Sc$ attains for constant $Y^0$ a local minimum at the conifold point. There exists, however, also a local maximum around $T\approx 0.05$ (Fig.~\ref{fig:nonsusyS}). As mentioned at the end of the previous section, this point is an attractor,
provided that one can still trust the prepotential \eqref{prepot} there.
If we keep $Y^0$ constant as in \cite{Gukov:2005bg}, the maximal value of 
the non-BPS entropy is higher than that of the BPS entropy.  Therefore,
the corresponding non-supersymmetric flux vacuum is entropically favored.

\section*{Acknowledgements}

We would like to thank B.~de Wit and S. Mahapatra
for valuable discussions.
This work is partly supported by EU contract MRTN-CT-2004-005104.

\phantomsection
\addcontentsline{toc}{section}{\numberline{}References}
\bibliographystyle{JHEP-3}
\bibliography{bibliography}

\end{document}